\documentclass[journal]{IEEEtran}
\usepackage{latexsym}
\usepackage{amssymb}
\usepackage{amsmath} 
\usepackage{booktabs}
\usepackage{enumerate}
\usepackage{graphicx}
\usepackage{subfigure}
\usepackage{xspace}
\usepackage{float}
\usepackage{bbm}
\usepackage{bm}
\usepackage{multirow}
\usepackage{booktabs}
\usepackage{color}
\usepackage{framed}
\usepackage{stfloats}
\usepackage{iitem}
\usepackage{amssymb}
\usepackage{pifont}

\ifCLASSINFOpdf
\else
\fi
\begin{document}
%
\title{Dynamic Contrastive Distillation \\for Image-Text Retrieval}
%
%
%

\author{Jun~Rao$^*$,
        Liang~Ding,
        Shuhan~Qi$^\dag$,
        Meng Fang,
        Yang Liu,
        Li Shen,
        and~Dacheng~Tao,~\IEEEmembership{Fellow,~IEEE}
\thanks{J. Rao, S. Qi, and Y. Liu are with the Harbin Institute of Technology, Shenzhen, China; L. Ding, L. Shen and D. Tao are with the JD Explore Academy at JD.com, Beijing, China; M. Fang is with the Eindhoven University of Technology, Eindhoven, Netherlands.}
\thanks{$*$~Work was done when Jun was interning at JD Explore Academy.}
\thanks{$\dag$~Corresponding Author: shuhanqi@cs.hitsz.edu.cn}
}

%
%

\markboth{Journal of \LaTeX\ Class Files,~Vol.~14, No.~8, August~2015}%
{Shell \MakeLowercase{\textit{et al.}}: Bare Demo of IEEEtran.cls for IEEE Journals}
%



\maketitle

\begin{abstract}
Although the vision-and-language pretraining (VLP) equipped cross-modal image-text retrieval (ITR) has achieved remarkable progress in the past two years, it suffers from a major drawback: the ever-increasing size of VLP models restrict its deployment to real-world search scenarios (where the high latency is unacceptable).
To alleviate this problem, we present a novel plug-in dynamic contrastive distillation (DCD) framework to compress the large VLP models for the ITR task.
Technically, we face the following two challenges:
1) the typical uni-modal metric learning approach is difficult to directly apply to cross-modal task, due to the limited GPU memory to optimize too many negative samples during handling cross-modal fusion features.
2) it is inefficient to static optimize the student network from different hard samples, which have different effects on distillation learning and student network optimization. 

We try to overcome these challenges from two points. First, to achieve multi-modal contrastive learning, and balance the training costs and effects, we propose to use a teacher network to estimate the difficult samples for students, making the students absorb the powerful knowledge from pre-trained teachers, and master the knowledge from hard samples.
Second, to dynamic learn from hard sample pairs, we propose dynamic distillation to dynamically learn samples of different difficulties, from the perspective of better balancing the difficulty of knowledge and students' self-learning ability.

We successfully apply our proposed DCD strategy on two state-of-the-art vision-language pretrained models, i.e. ViLT and METER.
Extensive experiments on MS-COCO and Flickr30K benchmarks show the effectiveness and efficiency of our DCD framework. Encouragingly, we can speed up the inference at least 129$\times$ compared to the existing ITR models.
We further provide in-depth analyses and discussions that explain where the performance improvement comes from. We hope our work can shed light on other tasks that require distillation and contrastive learning.
The code will be released upon acceptance.
\end{abstract}


\IEEEpeerreviewmaketitle

\section{Introduction}
\begin{figure}[t!]
\centering
\subfigure[Original knowledge distillation framework]{
\begin{minipage}[t]{0.8\columnwidth}
\centering
\includegraphics[width=\columnwidth]{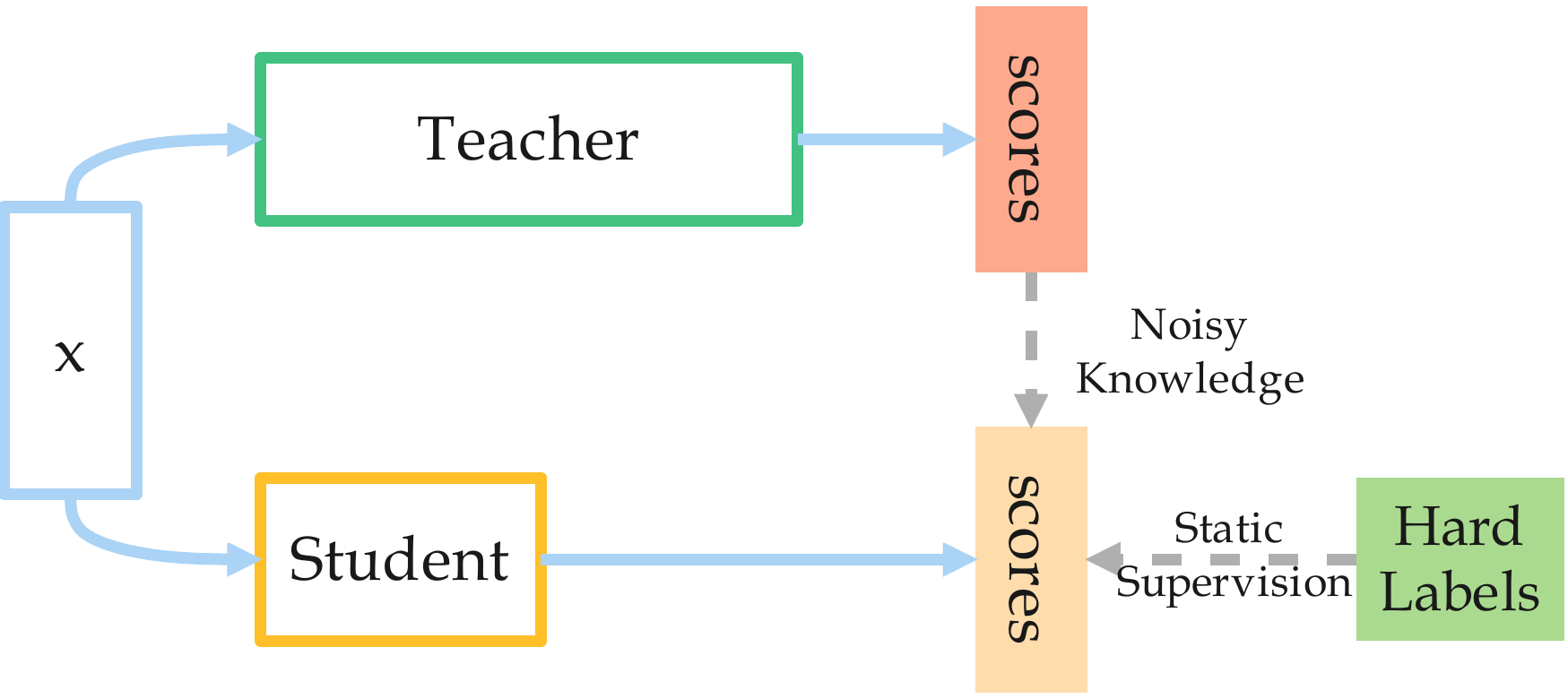}\label{1a}
\end{minipage}%
}%
\\
\subfigure[Our dynamic contrastive distillation framework]{
\begin{minipage}[t]{0.9\columnwidth}
\centering
\includegraphics[width=\columnwidth]{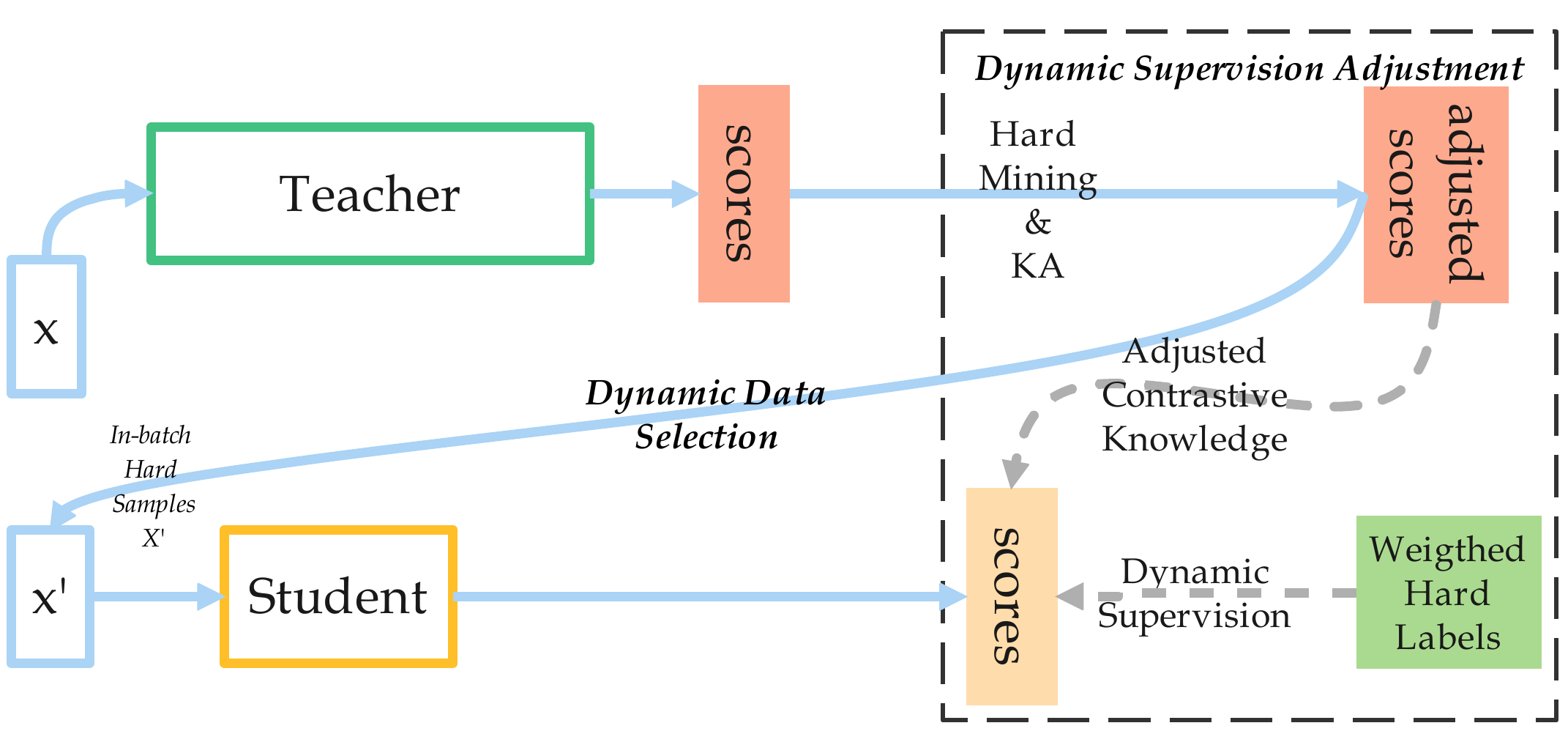}\label{1b}
\end{minipage}%
}%

\caption{Comparison of an existing distillation method and proposed framework. (a) The original distillation method supervises students through hard labels and uncorrected soft labels ``Noisy Knowledge''.  (b) 
In our DCD framework, we use the structure of dynamic data selection (\S \ref{sec:dd}) and dynamic supervision adjustment (\S \ref{dsa}) to get in-batch hard samples and weighted labels to supervise students dynamically with adjusted scores and weighted hard labels.
We also combine knowledge adjustment abbreviated as KA (\S \ref{ka}) to get adjusted contrastive knowledge.
}
\label{fig:introduction}
\end{figure}
\IEEEPARstart{M}{ultimodal} learning becomes a surging topic due to the increasing accessibility of multimodal data, such as image, text, video and audio. Also, with the advances of hardware, neural network models are able to scale up their capacity and expressive power to better leverage information from multiple modalities\cite{DBLP:journals/tmm/KyawQGZZXWC17}\cite{DBLP:journals/tmm/LuLNCZ21}.

Multimodal learning  (e.g. cross-modal retrieval) becomes a popular research topic~\cite{DBLP:journals/tmm/LiYLNX19}.
Image-text retrieval focuses on obtaining a set of sentences given a query image, namely measuring cross-modal similarity of image and text.
How to accurately measure the similarity of images and texts is at the core of this task.
Most image-text retrieval methods adopt the idea of contrastive learning~\cite{ DBLP:journals/corr/abs-1807-03748/Contrastive,ye2019unsupervised,caron2021emerging}. 
In this paradigm, different modalities are encoded into a semantic space to obtain modal-agnostic semantic representation, such that
the representations of a query and its corresponding matching key are clustered while unmatched key-query pairs are separate.

Constructing a dynamic dictionary might be considered a typical contrastive learning strategy~\cite{DBLP:conf/cvpr/He0WXG20/moco,DBLP:conf/icml/ChenK0H20/simclr,DBLP:conf/nips/GrillSATRBDPGAP20/byol}. 
In the dictionary, each word is a sample embedding encoded by the network.
Essentially, such a dynamic dictionary can be regarded as a global negative sample pool to explore informative sample pairs. When the dictionary is large enough and contains enough negative samples, the encoder can extract more discriminative features.
In this way, large and consistent dictionaries can be constructed for unsupervised learning with a contrastive loss~\cite{DBLP:conf/cvpr/HadsellCL06/contrast_learning}.

Knowledge distillation (KD) \cite{DBLP:journals/corr/HintonVD15} is to obtain a much smaller model with comparable performance, while greatly reducing the memory usage and accelerating the model inference. It has been widely used in recent years in natural language processing (NLP) and computer vision (CV) tasks \cite{tinybert,PKD,rco,Knowledge_Review}. 
However, when applying KD for image-text retrieval, there are incompatibilities between the contrastive learning paradigm and image-text retrieval~\cite{SCAN,uniter,vilbert}.

On one hand, the dictionary \cite{DBLP:conf/cvpr/He0WXG20/moco} is not suitable for image-text retrieval learning. 
To acquire multimodal content embedding, most multimodal models must interact with distinct changing modalities of embedding.
As a result, the slowly updated dictionary cannot maintain a large and diversified sample pool.

On the other hand, the existing image-text retrieval systems are too inefficient to learn the separability of sample pairs~\cite{vse++,uniter,vilbert}. 
Though state-of-the-art (SOTA) methods, e.g. UNITER~\cite{uniter} and ViLBERT~\cite{vilbert} using self-distillation~\cite{DBLP:conf/iccv/ZhangSGCBM19}, utilize an intermediate model to take a large number of random samples and select the top $K$ sample pairs that are most similar to fine-tune the image-text retrieval model, it is inefficient due to the lack of a stable selection of hard negative samples and appropriate information guidance, as well as the fact that it is computationally intensive for contrastive learning. 
Because of the limited information and low gradient values of these randomly picked samples, their contributions to the training may be less informative since many of them already satisfy the loss's requirements. 
Besides, a longer training time is necessary to make the network converge. 
These problems lead to our first research question (RQ): \\
\textit{RQ1: How to facilitate the contrastive learning paradigm in the distillation of image-text retrieval with informative samples?} 

Different from those recent SOTA methods \cite{uniter,vilbert}, we directly introduce a teacher network to train a smaller student using knowledge distillation (KD)~\cite{DBLP:journals/corr/HintonVD15} to more efficiently learn the differences and similarities of samples within constrained resources, as shown in Figure \ref{1a}. 
This teacher network may convey knowledge to a student network and select informative samples for training the multimodal interaction layers. 
A well-learned teacher network is flexible in selecting hard samples, and stabilizes the training process.
To make an analogy with the real world, we equate teacher networks with professors and student networks with graduate students. 
Professors typically have a certain level of knowledge and are aware of which topics are currently challenging, and it is advantageous for students to follow these topics.

Another issue with visual language model compression is
how to make greater use of the available information of teacher.
However, the vanilla KD approaches as shown in Figure \ref{1a} are static and unable to actively learn from different samples, consequently failing to learn the teacher's separability of sample pairs well. Another research question arises as a result of this: \\
\textit{RQ2: For the purpose of improving teacher knowledge transfer, is it possible to use the weighting method to dynamically learn diversified content,  according to the information of the limited sample pairs?}

We design a basic weighting strategy based on the teacher's uncertainty of samples and achieve dynamic distillation by adjusting the sample's contribution to the training. The core of our solution is that, for the distillation loss item, we pay particular attention to the samples that the teacher believes to be mastered. Concerning the task loss item, we place a higher weight on samples that the teacher believes to be confusing so that students can focus on learning these samples through the hard label.

As shown in Figure \ref{1b}, our framework addresses the two research questions described above.  Specifically, we first filter out the more informative and difficult samples for students to learn from the samples selected by the teacher. Second, we obtain adjusted scores by knowledge adjustment (KA) and weigh these samples by teacher uncertainty as soft labels for supervising student networks. Finally, we use teacher uncertainty to weigh the hard labels to obtain dynamic supervision signals to enhance students' self-learning ability. In this paper, we validate our approach in different training settings and benchmark datasets upon a single cross-modal fusion layer based on ViLT \cite{kim2021vilt} model.
The experimental results indicate that dynamic data selection and supervision weighting improve image-text retrieval performance. 
At the same time, we use METER~\cite{dou2022an} with a different architecture using co-attention \cite{vilbert} to achieve the same promising effects.
Our contributions are:
\begin{itemize}
    \item We propose to leverage the teachers' adjusted knowledge to mine in-batch hard samples and supervise the students (\S \ref{sec:dd});
    \item We explore a variety of sample level weighted settings to achieve better teacher knowledge transformation (\S \ref{dsa});
    \item Considering above aspects, we design a plug-in DCD framework to compress VLP models and guarantee competitive results compared to SOTA distillation approaches (\S \ref{exp});
    \item To the best of our knowledge, we are the first to dynamically distill a pre-trained model based on Transformer architecture with modalities' interaction for image-text retrieval.
\end{itemize}

\section{Related work}
\begin{figure*}[t!]
    \centering
    \includegraphics[width=0.9\linewidth]{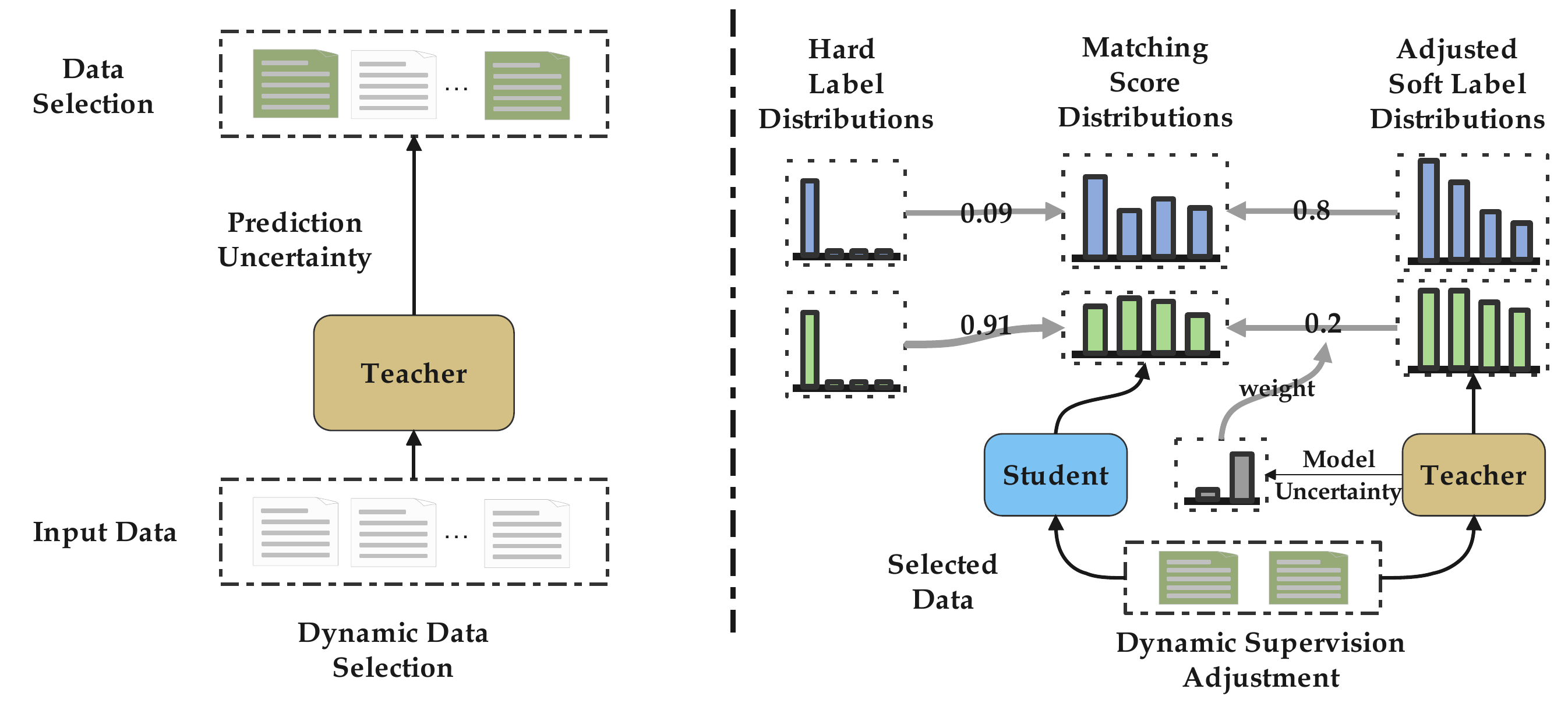}
    \caption{Illustration of our dynamic contrastive distillation framework (DCD) in two aspects: data selection(\textit{left}) and supervision adjustment(\textit{right}). We obtain more informative data samples through the uncertainty of the teacher network, and use the uncertainty to weight the degree of supervision of the soft and hard labels.} 
    \label{fig:overview}
\end{figure*}

\subsection{Image-Text Retrieval}
It is difficult to represent and match the semantic information of many modalities in image-text retrieval.
Recently, numerous existing methods \cite{vse++,DBLP:conf/iccv/LiZLLF19/VSRN,DBLP:conf/mm/WuWSH19/saem,DBLP:conf/mm/Qu0CN020/campera,DBLP:journals/tmm/HeXKWP16} for image-text retrieval encode the features into a semantic space using modality-independent encoders and then perform modal fusion to obtain the corresponding fusion features for cross-modal matching.

Some methods~\cite{DBLP:conf/mm/WuWSH19/saem,vse++} investigate self-attention to improve the feature embedding of intra-modality and then measure distance in a common metric space to use contrastive learning. In practical application scenarios, this type of approach usually allows the encoding of each modality to be calculated in advance, and only involves the calculation of the dot product of each modal vector at the time of retrieval. Thus such an approach is usually more flexible for large-scale retrieval, while these pre-encoded feature vectors of individual modes can be used for other downstream tasks. However, this type of method usually brings worse reproducibility stability and poorer performance \cite{ding2022repro}.

Others \cite{SCAN,Wang_2019_ICCV/camp,kim2021vilt,vilbert,uniter} adopt  inter-modality interaction to obtain a robust multi-model representation. With a sophisticated cross-modal attention mechanism or a graph neural network, such methods are able to achieve state of the art in cross-modal tasks. These work demonstrate the importance of multi-modal interaction layers. 
Existing SOTA methods \cite{uniter}~\cite{dou2022an} for ITR have used the transformer architecture for modal interaction between texts and images to obtain multi-modal fusion features. However, this interaction increase the complexity for retrieval  (the fusion of the embedding vectors of each modality and too large amount of parameters) and not suitable in practical applications (long inference time). Specifically, if the two modalities have $m$ and $n$ samples, respectively, the complexity of such fusion often has a greater complexity ($O(mn)$), compared to the modality-independent approach~\cite{vse++,DBLP:conf/mm/WuWSH19/saem} $(O(m+n))$.

Therefore we focus on the latter type of approach. 
If such a heavy modal interaction layer can be compressed and obtained close to the original model, our training time, inference time can be greatly reduced and applied to similarly structured ITR models, which are more practical in real-world deployment.

\subsection{Knowledge Distillation}

KD \cite{DBLP:journals/corr/HintonVD15} is presently one of the most appealing methods for compression in BERT-like models \cite{bert} and can be applied in ITR for compressing the multi-modal interaction layers. 

The theory behind KD is that a large teacher model can teach a small student model to imitate the teacher's behavior. In this way, the knowledge contained in the teacher model can be effectively transferred to the student model.
A set of methods~\cite{distilbert,PKD,tinybert,mobilebert} use intermediate state matching and logit matching to distill a pre-trained language model for downstream tasks, achieving a strong compression effect while performing almost the same as the original model. Nevertheless, these methods still need to use a large-scale unlabeled corpus for distillation, which requires a lot of computational resources~\cite{DBLP:journals/corr/abs-2205-15308}.
In the multimodal field, there are also several works \cite{Fang_2021_ICCV,DBLP:conf/cikm/RaoQQW0021,DBLP:journals/corr/abs-2104-13921} that introduce KD to compress visual-and-language (VLP) models. Since the baseline models used in these methods are similar to BERT, most of them migrated to the BERT compression method of NLP and obtained good distillation results. After a simple application of KD, only a few works \cite{DBLP:conf/emnlp/LiLRLZ021,DBLP:conf/bmvc/TangFSKZL19,DBLP:conf/eccv/ZhangLDZBCW20,fang2017learning} explore how to further transfer the teacher's knowledge to the student. They follow the idea of active learning to choose the data and the degree of learning for each sample. 

Inspired by above works, we use ViLT \cite{kim2021vilt} and METER~\cite{dou2022an}, the powerful pre-trained models of the BERT family, and investigate how to perform dynamic and adaptive distillation to achieve better distillation results in multimodal retrieval tasks. Meanwhile, different from the recent weighting methods \cite{DBLP:conf/iccv/LinGGHD17/focalloss,DBLP:conf/eccv/ZhangLDZBCW20}, we introduce the teacher-student framework to obtain the uncertainty score by a well-known teacher network rather than the student model itself.
\subsection{In-batch Hard Sample Mining}\label{hsm}
Although our distillation framework is consistent with the goal in \cite{DBLP:conf/nips/Sohn16} to obtain more informative negative samples to optimize learning, our implementation is quite different from \cite{DBLP:conf/nips/Sohn16} and is more applicable to multi-modal settings.

The key to contrastive learning is using a small number of samples to approximate the distribution of the data. Therefore, in-batch hard sample mining to better utilize the informative negative samples has been extensively studied~\cite{DBLP:conf/iccv/LinGGHD17/focalloss,li2020unimo,uniter,DBLP:conf/cvpr/ShrivastavaGG16,DBLP:conf/aaai/Li0W19}.
One of the mainstream methods for in-batch hard sample mining is online mining, which uses the loss ~\cite{DBLP:conf/cvpr/ShrivastavaGG16} or different gradients ~\cite{DBLP:conf/aaai/Li0W19} of samples in a batch to decide whether samples are hard.
For most image-text retrieval methods~\cite{vse++,SCAN,DBLP:conf/cvpr/caan}, the negative samples with the highest similarity in a batch are selected online as the informative in-batch hard negative samples so that there is no need to calculate other negative samples. 
The SOTA models, ViLBERT \cite{vilbert} and UNITER \cite{uniter} adopt checkpoints at different stages of training to select in-batch hard samples, which help it build diverse sample pools. However, such methods scarifice time and resources to achieve better performance.

In contrast, we achieve the same goal (mining informative hard negatives) in a different way through data selection and knowledge adjustment while striking a balance between time and performance (\S \ref{sec:dd}). 

\section{Preliminary}
\subsection{Contrastive Learning}\label{sec:contranstive}
In contrastive learning \cite{DBLP:conf/cvpr/HadsellCL06/contrast_learning,Wang2022ACC}, a representation space is commonly obtained by mixing training with positive and negative samples. Though training, the semantically similar samples are closer together in this space, whereas the semantically dissimilar samples are separated from each other.
Unlike typical self-supervised contrastive learning~\cite{DBLP:conf/cvpr/He0WXG20/moco,Zhong2022E2S2ES}, image-text retrieval uses a supervised learning paradigm.
In general, for a query image $v_i$, the matching sentence $t_i$ can be obtained directly according to the data annotation, and the unmatched sentence $t_j$ is obtained by randomly sampling non-labeled related sentences.
Given a set of image-text pairs $\{(q_i,k_i)\}{^N_{i=1}}$, our goal here is to use a contrastive learning approach to learn an optimal scoring function such that the scores of the matched image-text pairs ${(q_i,k^{+}_i)}$ are higher than the scores of the rest of the unmatched samples ${(q_i,k^{-}_j),j\neq i}$. 

From the probabilistic perspective, aligning $k_i$ to $q_i$ is equivalent to maximizing the conditional probability $p(k_i|q_i)$ while minimizing the probability for all negative pairs $p(k^{-}_j|q_i), j \neq i$. 
According to \cite{DBLP:journals/jmlr/GutmannH10/nceLOSS}, $p(k_j|q_i)$ can be approximated as:
\begin{equation} \label{eq:p_distribute}
p\left(k_{j} \mid q_{i}\right) \sim \frac{\exp ^{s\left(q_{i}, k_{j}\right)}}{\sum_{m=1}^{N} \exp ^{s\left(q_{i}, k_{m}\right)}}
\end{equation}
where $s(q_i, k_j)$ is the matching score between $q_i$ and $k_j$; the denominator is a sum over all possible sentences, which is a partition function for normalization. Therefore, NCE loss \cite{DBLP:journals/jmlr/GutmannH10/nceLOSS} can be measured in a softmax fashion:
\begin{equation}
\label{eq:nce}
\begin{split}
\mathcal{L}_{NCE} &=\sum_{i=1}^{N}-\log p\left(k_{i} \mid q_{i}\right)\\
& \sim \sum_{i=1}^{N}-\log \left(\frac{\exp ^{s\left(q_{i}, k_{i}\right)}}{\exp ^{s\left(q_{i}, k_{i}\right)}+\sum_{m \neq i} \exp ^{s\left(q_{i}, k_{m}\right)}}\right)
\end{split}
\end{equation}
The denominator in Equation \ref{eq:nce} requires a sum over all sentences in a dataset, which is intractable in practice. Therefore, we usually compute the NCE loss on a mini-batch of $K(K \ll N)$ image-text pairs sampled from the whole dataset.

\subsection{Knowledge Distillation} \label{section:kd}
For vanilla KD~\cite{DBLP:journals/corr/HintonVD15}, we need a teacher network to guide the student network. We consider a single mini-batch and let $\boldsymbol{z^k_i}$ be the $k$-th value of the logit vector $\boldsymbol{z_i}$. The initial teacher and student model can be defined as: teacher $\boldsymbol{p}(\theta^t)$ and student $\boldsymbol{p}(\theta^s)$, respectively, where $\theta$ is the net parameters and $\boldsymbol{p^k}(\cdot)=\frac{\exp \left(\boldsymbol{z}^{k}(\theta) / \tau\right)}{\sum_{j=1}^{K} \exp \left(\boldsymbol{z}^{j}(\theta) / \tau\right)}$ is the probability predict of the matching label and $K$ is the number of classes. So the KL divergence distillation loss can be defined as:
\begin{equation}\label{eq6}
\mathcal{L}_{K L}\left(\boldsymbol{p}(\tau|\theta^s), \boldsymbol{p}(\tau|\theta^t)\right)=\tau^{2} \sum_{j} \boldsymbol{p}^{j}(\tau|\theta^t)\cdot\\\log \frac{\boldsymbol{p}^{j}(\tau|\theta^t)}{\boldsymbol{p}_{j}(\tau|\theta^s)}
\end{equation}
where $\tau$ is the temperature factor used in KD, which controls how much to rely on the teacher's soft predictions. For simplicity of notation, we use $\mathcal{L}_{K L}$ to represent $\mathcal{L}_{K L}\left(\boldsymbol{p}(\tau|\theta^s), \boldsymbol{p}(\tau|\theta^t)\right)$.

For a better distillation effect, we follow \cite{DBLP:conf/ijcai/KimOKCY21} and use Mean Squared Error (MSE) loss. The MSE Loss can be defined as follows:
\begin{equation}
   \mathcal{L}_{MSE} = || \boldsymbol{z}(\theta^s) - \boldsymbol{z}(\theta^t) ||_2^2
\end{equation}
We can therefore get the final loss $\mathcal{L}$ of the student network:
\begin{equation}\label{eq:base_KD_loss}
\begin{gathered}
    \mathcal{L} = \alpha\mathcal{L}_{MSE} +(1-\alpha) \mathcal{L}_{task}
\end{gathered}
\end{equation}
where $\alpha$ is the hyper-parameter that balances the importance of the task loss $\mathcal L_{task}$ and the distillation loss $\mathcal L_{MSE}$.
\section{Method}\label{method}
Figure~\ref{fig:overview} illustrates our \textsc{dynamic contrastive distillation} framework in two aspects. From the data side (``Dynamic Data Selection''), we select the informative samples for students according to the teacher's uncertainty estimation. From the supervision side (``Dynamic Supervision Adjustment''), we use teacher uncertainty to select the level of importance of supervision.

\subsection{Dynamic Data Selection and Knowledge Adjustment}\label{sec:dd}
\subsubsection{Dynamic Data Selection}\label{section:HNM}
Contrastive learning~\cite{DBLP:conf/cvpr/He0WXG20/moco} benefits from a large batch size~\cite{DBLP:conf/icml/ChenK0H20/simclr} and extensive data augmentation~\cite{DBLP:journals/corr/abs-2112-09331/zerovl}.
However, the high computational cost of multimodal fusion layers hinders its wide usage.

As defined in Section \ref{sec:contranstive}, $\mathcal{L}_{NCE}$ requires the network to pass all $N \times N$ pairs into the multimodal layers. 
Assuming that the number of tokens in the image is $m$, the number of tokens in the text is $n$, and the dimension of each token is $d$, then the complexity of the self-attention mechanism is $O\left(d(m+n)^{2}\right)$. Due to such high computational complexity, most image-text retrieval methods adopt a smaller value of $K(K \ll N)$. NCE loss directly samples positive samples in a mini-batch of $K$ pairs and get the remaining $K \times (K-1)$ mismatched pairs. 
In typical methods \cite{uniter,vilbert,SCAN}, they take only one negative sample. Thus Equation \ref{eq:nce} becomes the following:
\begin{equation}\label{eq:itm}
\begin{split}
\mathcal{L}_{ITM} &= \sum_{i=1}^{K}-\log \left(\frac{\exp ^{s\left(q_{i}, k^{+}_{i}\right)}}{\exp ^{s\left(q_{i}, k^{+}_{i}\right)}+ \exp ^{s\left(q_{i}, k^{-}_{i}\right)}}\right)
\end{split}
\end{equation}
where $k^{+}$ is the paired key of the corresponding query, and $k^{-}$ is the unmatched negative key. In particular, most methods usually use the way of VSE++ \cite{vse++} to calculate more negative samples, and only update the gradient of the hardest negative sample during back propagation.

Motivated by~\cite{DBLP:conf/iclr/RobinsonCSJ21} and \cite{li2020unimo}, we use the teacher's information to pick in-batch hard examples to feed into the student model for supervised learning.

We get fewer but more useful negative samples by taking larger random negative samples and propagating them through the teacher network.
As shown in Figure \ref{1b}, we begin by increasing the number of in-batch negative samples obtained by the teacher network to $M$. Then we calculate $M$ negative sample pairs and a positive sample pair in the teacher network to obtain the logits over the binary class of $M+1$ matching scores. Then we sort the scores, take the $M'+1$ largest scores, and find the corresponding samples and their matching scores. We input these informative samples into the student network, and get the logits of the $M'+1$ matching scores of the student network. The new image-text matching loss function can be defined as follows:
\begin{equation}\label{eq:itm'}
\begin{split}
\mathcal{L}_{ITM^{'}} &= \sum_{i=1}^{K}-\log \left(\frac{\exp ^{s\left(q_{i}, k^{+}_{i}\right)}}{\exp ^{s\left(q_{i}, k^{+}_{i}\right)}+ \sum_{j=1}^{M^{'}}\exp ^{s\left(q_{i}, k^{-}_{ij}\right)}}\right)
\end{split}
\end{equation}
Then we use the teacher's logits score to constrain the student network. In this way, we not only reduce the gradient calculation of the student network but also improve the ability of the student network. Naturally, if a larger in-batch negative sample value ($M$) is adopted, the performance of the network will be improved to a certain extent, but at the same time, it will increase the network learning time. We need to adjust this value ($M$) and the number of negative samples ($M'$) that students need to learn in practical applications.

\subsubsection{Knowledge Adjustment}\label{ka}
In KD and KD-based approaches, the student network is trained under the supervision of teacher predictions, regardless of whether this supervision signal is right or wrong.
We select the indistinguishable and hard negative samples according to the teacher network.  Therefore, the scores of these negative samples may be higher than those of positive samples. We call this situation a ``genetic error''. At the same time, if a student continues to learn this erroneous knowledge under the supervision of the teacher, it will further lead to errors in the student network. Therefore, we try to fix the samples of this mini-batch where the teacher's prediction does not match the true label, which we call knowledge adjustment.

For simplicity, we consider a mini-batch with only one matching pair. As described in the aforementioned section, we obtained the matching scores of $M^{'} + 1$ samples with the teacher network. The matching scores are sorted in descending order, but the positive samples are not necessarily among the $M^{'} + 1$ samples. 
Based on this consideration, we move the positive sample to the top of the original matching score list to generate a new mapping of samples and matching scores. In this way, we ensure that the sample with the highest confidence score must be a positive sample.

When computing the loss, for implementation convenience, for a batch of positive and negative samples (e.g., one positive sample pair and 15 negative sample pairs), we set the first position in the sample list to be the matching positive sample pair and the remaining positions to be the negative sample pairs. In general, we expect the first score to be the highest (because it is a positive sample pair), so we guarantee the nature of the highest matching score for the positive sample pair by performing such an insertion operation on the list of scores inferred by the teacher model, which can correct some erroneous outputs.
This method also keeps the numerical distribution of soft targets, which is helpful in stabilizing the training process.

With such a simple implementation of dynamic data selection and knowledge adjustment, we reduce the computational cost of training as well as bring a more performance and inference time balanced student model, and also bring a simple and effective alternative for practical lightweight ITR model deployment.
\subsection{Dynamic Supervision Adjustment}\label{dsa}
In the actual data annotation for image-text retrieval, we can know that image and text matching (1) or not matching (0), so we call this label information as “hard labels”. We can also get  the output of the teacher, which represents the degree of matching (normalized to a value between 0 and 1). This time the more matching samples are closer to 1 for the output of matching scores, and the more not matching samples scores are closer to 0. The output is a score to indicate how an image and a text match. We call this score as “soft labels”. In Hindon \cite{DBLP:journals/corr/HintonVD15}, it shows the information of soft labels is easier to learn compared to hard labels because of the inclusion of inter-class differences. Similarly, in image-text retrieval, soft labels are more representative of how well different sample pairs match, and bring more information worth learning.

We select more valuable learning samples through the teacher network and ensure the relative correctness of the predictions of the teacher, but there are still uncertain samples. Intuitively, for samples with high degree of certainty considered by the teacher, the soft labels provided by the teachers bring more significant learning information than the hard labels. 
Although these samples can provide a certain amount of information, if the student network completely relies on the guidance of the teacher's judgement at this time, this output may mislead feature learning in the fine-tuning stage and hurt adaptation performance.

We aim to reduce the negative influence of noisy soft labels by evaluating the credibility of these soft labels for each sample and reweighting the contributions of samples with error-prone predictions in the NCE loss and KD loss. In order to improve learning on such samples, we divide them into two parts.
\subsubsection{Weighted Hard Labels}
We use entropy to weight the hard labels in the task loss term, similar to previous works \cite{DBLP:conf/iccv/LinGGHD17/focalloss,DBLP:conf/emnlp/LiLRLZ021} that assign sample-wise weights.
In contrast to \cite{DBLP:conf/bmvc/TangFSKZL19}, we focus on self-exploration of task loss and reduce attention to the most difficult samples. 
Intuitively, the greater the teacher's uncertainty about the output of a sample, the lower the reliability of the output.
Therefore, we improve the student model of self-exploration by increasing its attention to the sample of task loss in which the teacher is highly uncertain.

Given $N$ instances in one batch, the corresponding output matching scores probability distribution of the student model over the positive-and-negative pair index $y$ (position 0 is the positive pair, and the others are negative pairs) is $p (y \mid x_i) $ like Equation~\ref{eq:p_distribute}, denoting the model confidence towards the positive pair.
The uncertainty score $u$ of teacher about the output of ${x_i}$ can be defined as Equation \ref{eq:uncertainty} with negligible computational overhead:
\begin{equation}\label{eq:uncertainty}
    u_{x_i}=-\sum_{y} p(y \mid x_i) \log p(y \mid x_i)
\end{equation}
Then we normalized the uncertain results to get the weight of the corresponding sample.
\begin{equation}
    w_i= -\frac{u_{x_i}} {\sum _{i=1}^{K} u_{x_i}}
\end{equation}
Finally, we combine the previous ITM loss to get the final weighted task loss as shown in Equation~\ref{eq:witm}.
\begin{equation}\label{eq:witm}
    \mathcal{L}_{WITM} =-\sum_{i=1}^{K}w_i\log \left(\frac{\exp ^{s\left(q_{i}, k^{+}_{i}\right)}}{\exp ^{s\left(q_{i}, k^{+}_{i}\right)}+ \sum_{j=1}^{M^{'}}\exp ^{s\left(q_{i}, k^{-}_{ij}\right)}}\right)
\end{equation}

\subsubsection{Weighted Soft Labels}
Similar to \cite{DBLP:conf/eccv/ZhangLDZBCW20}, we found that reducing the weight of the in-batch hardest samples in the distillation term is beneficial to students' learning. 
It is possible that the teacher didn't generate the correct matching scores since these examples weren't well learned.
As a result, if we focus on these in-batch harder samples, the supervisory information provided by teachers will be limited and may not be correct, and such errors may propagate to the student.

Therefore, we reduce the soft label loss term to pay attention to such samples in order to re-learn the samples that the teacher failed to master in a relatively correct training direction for the student network. However, even if these more in-batch hard samples need to be weighted down, they can still provide some useful information for the student. Therefore, even if we reduce the weight of such samples, we should not use the \textsc{focal loss} style weights \cite{DBLP:conf/iccv/LinGGHD17/focalloss} to make the weight difference between these samples too large. As a result, we defined the following reversed weights $c$ based on the forward weights calculated by teacher entropy:
\begin{equation}
    c_i =\frac{ \exp ^{(1 - w_i)^2} }{\sum_{i=1}^{K} \exp ^ {(1 - w_i)^2}}
\end{equation}
Therefore, we combine the previous MSE loss to get the final weighted distillation loss as shown in Equation~\ref{eq:wds}:
\begin{equation}\label{eq:wds}
    \mathcal{L}_{WDS} = \sum_{i=1}^{K}c_i|| \boldsymbol{z_i}(\theta^s) - \boldsymbol{z_i}(\theta^t) ||_2^2   
\end{equation}

\subsection{Overall Learning Objective}
The training objective in our method is finding the optimal
$\theta^s$ by minimizing the combination of the above two weighted losses:

\begin{equation}\label{eq:new_KD_loss}
\begin{gathered}
    \mathcal{L}^{\prime} =\alpha \mathcal{L}_{WDS}+ (1-\alpha) \mathcal{L}_{WITM}
\end{gathered}
\end{equation}

\section{EXPERIMENTS}\label{exp}
\subsection{Datasets}
We conducted experiments on two widely-used benchmarks: MS-COCO~\cite{DBLP:conf/eccv/LinMBHPRDZ14} and Flickr30k~\cite{DBLP:journals/tacl/YoungLHH14}, which consist of 123,287 and 31,783 images, respectively, and each image has five corresponding sentence descriptions. 
We closely followed \cite{DBLP:journals/pami/KarpathyF17} to split the datasets. Concretely, the processed Flickr30k dataset contains 1,000, 1,000, and 29,783 images for testing, validation, and training, respectively. As for MS-COCO, 5,000 images for testing and 5,000 for validation, the rest 113,287 images are left for training.

During inference, the performance for image-text retrieval is reported by Recall at K (R@K), which represents the ranking proportion of queries with ground-truth within the top K. R@1, R@5 and R@10 are our evaluation metrics.

\subsection{Implementation Details}
We validated our proposed dynamic contrastive distillation on two state-of-the-art vision-language pretrained models, \textit{ViLT}~\cite{kim2021vilt} and \textit{METER}~\cite{dou2022an}, where ViLT is used for the main experiments (\S\ref{subsubsec:main}-\S\ref{subsubsec:ablation}) to demonstrate the effectiveness of our method, and METER is used in \S\ref{subsubsec:universality} to show the universality.
Specifically, in the main experiments, we use ViLT with 12 Transformer layers as a teacher model for all scenarios that require distillation. 
While we use a 6 layers of Transformer as the student network for both Flickr30k and MS-COCO, we leave their best settings in the original paper~\cite{kim2021vilt} as the default.
We compress the original model for 40 epochs and for 20 epochs on Flickr30k and MS-COCO datasets, respectively. 
We train our models on 8 SuperPoD NVIDIA A100 GPUs. 
Due to the limitation of computational resources and in order to achieve a better trade-off between training time and retrieval performance (see discussion in Section~\ref{subsubsec:data}), 
we set the number of negative samples to 63 and 7 for the teacher (i.e. $M$ in \S\ref{sec:dd}) and student (i.e. $M^{'}$ in \S\ref{sec:dd}), respectively, for dynamic data selection.

\subsection{Results and Analysis}
\begin{table*}[t!]
\small
\centering
\caption{Comparisons of existing methods experimental results on Flickr30K and MS-COCO test sets. ``*'' represents the results obtained from \cite{DBLP:conf/emnlp/ZhangHJIS20}, which removed the excessively time-consuming online in-batch hard sample mining process.}\label{over_compare}
\resizebox{1\textwidth}{!}
{
\begin{tabular}{cllcccccccccccc}
\hline
\multirow{3}{*}{Model} & \multicolumn{1}{c}{\multirow{3}{*}{Param}} & \multicolumn{1}{c}{\multirow{3}{*}{\begin{tabular}[c]{@{}c@{}}Time\\      (ms)\end{tabular}}} & \multicolumn{6}{c}{Flickr30K}                                            & \multicolumn{6}{c}{MS-COCO}                                              \\
                       & \multicolumn{1}{c}{}                       & \multicolumn{1}{c}{}                                                                          & \multicolumn{3}{c}{Text Retrieval} & \multicolumn{3}{c}{Image Retrieval} & \multicolumn{3}{c}{Text Retrieval} & \multicolumn{3}{c}{Image Retrieval} \\
                       & \multicolumn{1}{c}{}                       & \multicolumn{1}{c}{}                                                                          & R@1        & R@5       & R@10      & R@1        & R@5        & R@10      & R@1        & R@5       & R@10      & R@1        & R@5        & R@10      \\ \hline
SCAN \cite{SCAN}                  & $\sim$73M                                  & $\sim$900                                                                                     & 67.4       & 90.3      & 95.8      & 48.6       & 77.7       & 85.2      & 72.7       & 94.8      & 98.4      & 58.8       & 88.4       & 94.8      \\
CAMP \cite{Wang_2019_ICCV/camp}                   & $\sim$94M                                  & $\sim$1000                                                                                    & 68.1       & 89.7      & 95.2      & 51.5       & 77.1       & 85.3      & 72.3       & 94.8      & 98.3      & 58.5       & 87.9       & 95.0      \\
VSRN \cite{DBLP:conf/iccv/LiZLLF19/VSRN}                   & $\sim$204M                                 & -                                                                                             & 71.3       & 90.6      & 96.0      & 54.7       & 81.8       & 88.2      & 76.2       & 94.8      & 98.2      & 62.8       & 89.7       & 95.1      \\
SAEM \cite{DBLP:conf/mm/WuWSH19/saem}                   & $\sim$178M                                 & $\gg$900                                                                                        & 69.1       & 91.0      & 95.1      & 52.4       & 81.1       & 88.1      & 71.2       & 94.1      & 97.7      & 57.8       & 88.6       & 94.9      \\
ViLBERT-Base* \cite{vilbert}           & $\sim$285M                                 & $\sim$920                                                                                     & 76.8       & 93.7      & 97.6      & 59.1       & 85.7       & 92.0      & 77.0       & 94.1      & 97.2      & 62.3       & 89.5       & 95.0      \\
UNITER-Base* \cite{uniter}            & $\sim$174M                                 & $\sim$900                                                                                      & 78.3       & 93.3      & 96.5      & 62.9       & 87.2       & 92.7      & 74.4       & 93.9      & 97.1      & 60.7       & 88.0       & 93.8      \\
DCD (Ours)                   & $\sim$66M                                  & $\sim$7                                                                                       & 75.6       & 91.0      & 94.6      & 53.7       & 81.4       & 88.2      & 76.5       & 94.1      & 98.0      & 59.7       & 89.7       & 95.7      \\ \hline
\end{tabular}
}
\end{table*}

\begin{figure}[t!]
    \centering
    \includegraphics[width=0.9\linewidth]{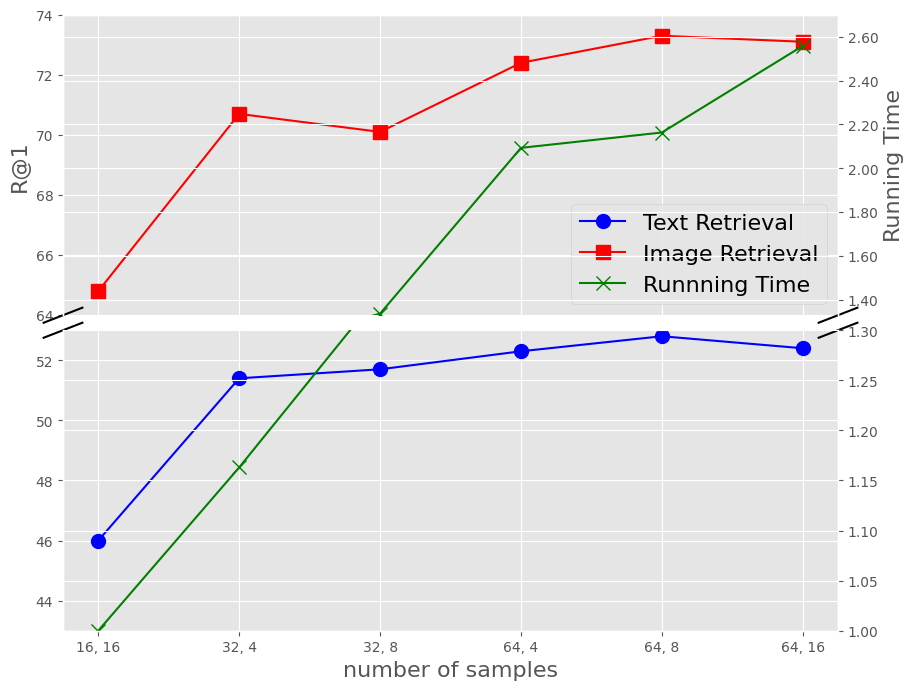}
    \caption{The effects on retrieval (``R@1'') and training time (``Running Time'') when inputting the different numbers of samples for teacher and student. ``(X, Y)'' in the horizontal coordinate indicates the numbers of samples randomly inputting to the teacher and student, respectively. The \textcolor{green}{green} line represents the corresponding running time of each setting, while the \textcolor{red}{red} and \textcolor{blue}{blue} lines show R@1 for text and image retrieval, respectively.
}
    \label{fig:negative_num}
\end{figure}
\begin{table*}[ht]
	\begin{center}
	\small
		\caption{Performance of the teacher and students with different loss re-weighting methods. ``*'' indicates that dynamic sample selection and knowledge adjustment are used. ``N/A'' means the training process does not converge, and it is almost impossible to retrieve correct results.}\label{dynamic_weight}
	\begin{tabular}{lccccc}
\hline
\multirow{2}{*}{Method} & \multicolumn{2}{c}{Flickr30K} & \multicolumn{2}{c}{MS-COCO} & \multirow{2}{*}{Avg.} \\
                        & TR@1          & IR@1          & TR@1         & IR@1         &                       \\ \hline
ViLT~\cite{kim2021vilt} (\textit{12L Teacher})                   & 83.7          & 62.2          & 83.7         & 68.4         & 74.5                  \\ \hline
\multicolumn{6}{c}{\textit{6L Student}}\\ \hline
Directly fine-tuning                & 63.3          & 42.0          & 68.6         & 47.5         & 55.4                  \\
Vanilla KD~\cite{DBLP:journals/corr/HintonVD15}              & 71.8          & 50.6          & 72.8         & 54.7         & 62.5                \\
WSL KD*~\cite{DBLP:conf/iclr/ZhouSCZWYZ21/wsl}                & 74.4          & 51.7          & 74.4         & 57.2         & 64.4                \\
Student-Uncertainy*      & N/A          & N/A          & N/A         & N/A        & N/A                  \\
DCD*                    & 75.6          & 53.7          & 76.5         & 59.7         & 66.4                \\ \hline
\end{tabular}
	\end{center}
\end{table*}
\begin{figure*}[t!]
\centering \label{figure:hnm}
\subfigure[The matching score of the \textbf{positive} pair.]{
\begin{minipage}[t]{\columnwidth}
\centering
\includegraphics[width=0.8\columnwidth]{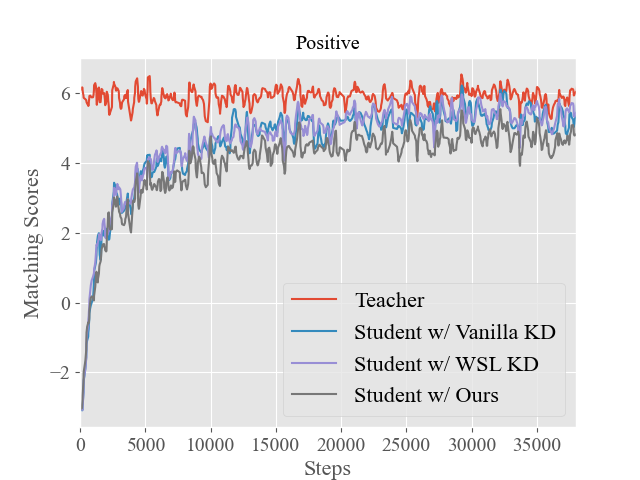}\label{4a}
\end{minipage}%
}%
\subfigure[The matching score of the \textbf{most similar negative} pair.]{
\begin{minipage}[t]{\columnwidth}
\centering
\includegraphics[width=0.8\columnwidth]{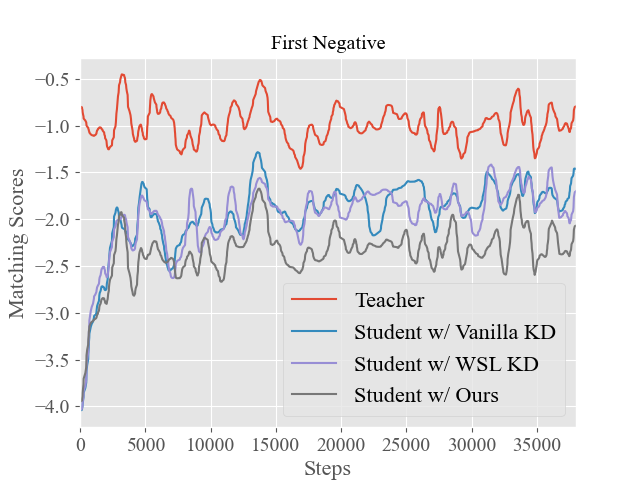}\label{4b}
\end{minipage}%
}%
\\
\subfigure[The matching score of the \textbf{second similar negative} pair.]{
\begin{minipage}[t]{\columnwidth}
\centering
\includegraphics[width=0.8\columnwidth]{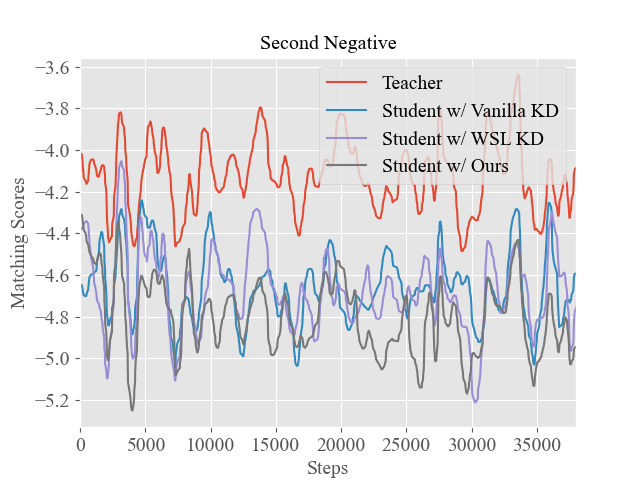}\label{4c}
\end{minipage}%
}%
\subfigure[The matching score of the \textbf{third similar negative} pair.]{
\begin{minipage}[t]{\columnwidth}
\centering
\includegraphics[width=0.8\columnwidth]{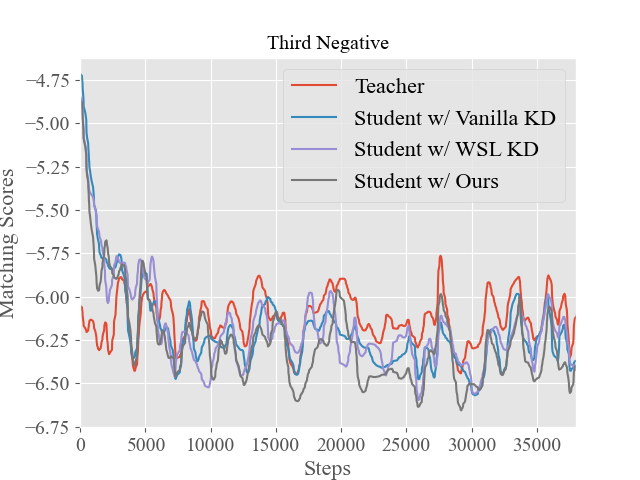}\label{4d}
\end{minipage}%
}%

\caption{Comparison of different dynamic supervisions, i.e. ``Student w/ \{Vanilla KD~\cite{DBLP:journals/corr/HintonVD15}/ WSL KD~\cite{DBLP:conf/iclr/ZhouSCZWYZ21/wsl}/ DCD (Ours)\}'' in terms of the matching score of true positives (a) and true negatives (b, c, d). The outputs of the teacher model are reported as the reference.
}
\label{fig:fig4}
\end{figure*}

\subsubsection{\textbf{Image-text Retrieval Results}}
\label{subsubsec:main}
In this section, we compare the two datasets of Flickr30K and MSCOCO to verify the effectiveness of our framework. Table~\ref{over_compare} shows the experimental results of Flickr30K and MSCOCO 1K test sets. 
For a comprehensive comparison, we list not only the \textit{retrieval results} (``R@K'') of the existing image-text retrieval methods, including SCAN~\cite{SCAN}, CAMP~\cite{Wang_2019_ICCV/camp}, VSRN~\cite{DBLP:conf/iccv/LiZLLF19/VSRN}, SAEM~\cite{DBLP:conf/mm/WuWSH19/saem}, ViLBERT~\cite{vilbert}, and UNITER~\cite{uniter}, but also their corresponding \textit{model sizes} (``Param'') and \textit{inference latency} (``Time'').
As shown in Table \ref{over_compare}, the inference speed of our compressed model (``Ours'') is significantly faster than that of the existing models, where the speedup is at least \textbf{129}$\times$ (7\textit{ms} vs. 900\textit{ms}). 
Also, our approach has the smallest model size among them, achieving up to 4.1$\times$ parameter compression (comparing to ViLBERT-Base).
Regarding retrieval performance, our compressed model achieves competitive, if not better, performance compared to the existing state-of-the-art in image-text retrieval.
Specifically, compared with similar-sized method \textsc{SCAN}~\cite{SCAN}, we achieve 8.2\% and 5.1\% improvements in R@1 text and image retrieval, respectively, in the Flickr30k dataset. In the MS-COCO 1K dataset, our method also obtains 3.8\% and 0.9\% improvement in text and image retrieval, respectively. 
Meanwhile, our compressed model gains consistent improvement compared to those larger non-pretrained modal interaction models, i.e. \textsc{CAMP}~\cite{Wang_2019_ICCV/camp}, \textsc{VSRN}~\cite{DBLP:conf/iccv/LiZLLF19/VSRN}, and \textsc{SAEM}~\cite{DBLP:conf/mm/WuWSH19/saem}. 
And encouragingly, compared to the state-of-the-art pretrained models, i.e. \textsc{ViLBERT}~\cite{vilbert} and \textsc{UNITER}~\cite{uniter}, our compressed model achieves better performance with significantly fewer parameters, such as TR@1, TR@10, and IR@10 on MS-COCO.
\begin{table}[ht]
	\begin{center}
	\caption{The impact of components in Flickr30K}\label{components_f}
		\begin{tabular}{lcccccc}
\hline
\multicolumn{1}{c}{\multirow{3}{*}{Method}} & \multicolumn{6}{c}{Flickr30K}                                            \\
\multicolumn{1}{c}{}                        & \multicolumn{3}{c}{Text Retrieval} & \multicolumn{3}{c}{Image Retrieval} \\
\multicolumn{1}{c}{}                        & R@1        & R@5       & R@10      & R@1        & R@5        & R@10      \\ \hline
Vanilla KD                                  & 71.8       & 90.3      & 94.1      & 50.6       & 79.8       & 87.6      \\
+DS\&KA                                     & 73.7       & 90.7      & 94.4      & 51.5       & 80.0       & 87.2      \\
\quad+HW                                        & 75.5       & 90.0      & 94.2      & 53.5       & 81.4       & 88.4      \\
\quad+SW                                        & 74.6       & 89.7      & 94.4      & 51.7       & 80.1       & 87.7      \\
+FULL                                       & 75.6       & 91.0      & 94.6      & 53.7       & 81.4       & 88.2      \\ \hline
\end{tabular}
	\end{center}
\end{table}

\begin{table}[ht]
	\begin{center}
	\caption{The impact of components in MS-COCO}\label{components_ms}
		\begin{tabular}{lcccccc}
\hline
\multirow{3}{*}{Method} & \multicolumn{6}{c}{MS-COCO}                                              \\
                        & \multicolumn{3}{c}{Text Retrieval} & \multicolumn{3}{c}{Image Retrieval} \\
                        & R@1        & R@5       & R@10      & R@1        & R@5        & R@10      \\ \hline
Vanilla KD              & 72.8       & 92.5      & 96.6      & 54.7       & 86.3       & 93.8      \\
+DS\&KA                 & 73.9       & 93.2      & 97.5      & 57.1       & 87.8       & 95.2      \\
\quad+HW                    & 74.5       & 93.4      & 97.5      & 58.1       & 87.6       & 94.7      \\
\quad+SW                    & 75.2       & 93.9      & 97.6      & 57.0       & 88.0       & 95.0      \\
+FULL                   & 76.5       & 94.1      & 98.0      & 59.7       & 89.7       & 95.7      \\ \hline
\end{tabular}
	\end{center}
\end{table}
\begin{table*}[ht]
	\begin{center}
	\small
	\caption{The Generalizability of DCD upon the SOTA VLP model -- METER~\cite{dou2022an}. ``*'' indicates that dynamic sample selection and knowledge adjustment are used.}\label{Generalizability}
\scalebox{0.96}{
\begin{tabular}{lcccccc}
\hline
\multirow{3}{*}{Method} & \multicolumn{6}{c}{Flickr30K}                                            \\
                        & \multicolumn{3}{c}{Text Retrieval} & \multicolumn{3}{c}{Image Retrieval} \\
                        & R@1        & R@5       & R@10      & R@1        & R@5        & R@10      \\ \hline
METER\cite{dou2022an}(\textit{6L Teacher})                   & 94.3      & 99.6     & 99.9     & 82.2      & 96.3      & 98.4     
               \\ \hline
\multicolumn{7}{c}{\textit{3L Student}}\\ \hline
Directly fine-tuning                  & 80.3      & 96.4     & 98.3     & 55.2      & 86.7      & 93.3     \\
Vanilla KD              & 92.7      & 99.4     & 99.8     & 79.8      & 96.4      & 98.4    \\
WSL KD*                 & 92.9      & 99.3     & 99.8     & 80.9      & 96.6      & 98.6     \\
DCD*                    & 93.4      & 99.3     & 99.8     & 81.9      & 96.6      & 98.7     \\ \hline
\end{tabular}}
	\end{center}
\end{table*}

\subsubsection{\textbf{Analysis of Dynamic Data Selection}} 
\label{subsubsec:data}
Recall that we denote the $M$ and $M'$ by the number of negative samples as input to the teacher and student network, respectively. 
These two factors may significantly influence both \textit{retrieval performance} (Recall) and \textit{training costs} (Time). 
To achieve the desired trade-off, we carefully investigate the effects of them spanning a reasonable range, that is, ($M$+1, $M'$+1)\footnote{``+1'' means our settings take one positive sample and the rest are negative samples.} $\in$ \{(16, 16), (32, 4), (32, 8), (64, 4), (64, 8), (64, 16)\} shown in Figure~\ref{fig:negative_num}. Note that the training costs of adding negative samples to the teacher network is substantially lower than that of the student network. We therefore can reduce the time cost of the student when mining difficult samples by increasing the sample input of the teacher network.

As seen in Figure \ref{fig:negative_num}, increasing the number of (negative) samples can basically obtain better text and image retrieval results (see the red and blue lines) but significantly enhance the training costs (see the green line), validating the effectiveness of negative samples in our dynamic contrastive distillation frameworks. We also show several interesting findings: \textit{1)} increasing the number of hard negative samples of students does not improve the retrieval if we set a relatively small number of negative samples for teachers, e.g. ($M$+1, $M'$+1) changes from (32, 4) to (32, 8), demonstrating the necessity of setting a relatively large number of negative samples for teachers; \textit{2)} increasing the number of negative samples for students causes the retrieval results to rise first and then decline. It shows that while there are some hard samples in the batch, it also increases the number of simple sample pairs, which hurts the network's final result. This is similar to what VSE++~\cite{vse++} reported. Based on observations, to achieve the desired trade-off, we set the number of negative samples to 63 and 7 (in total 64 and 8) for the teacher and student, respectively, for in-batch hard sample dynamic selection.

\subsubsection{\textbf{Analysis of Dynamic Supervision Adjustment}}
\label{subsubsec:supervision}
We first empirically show the superiority of dynamic supervision adjustment, then discuss where the improvement comes from?

\paragraph{\textbf{The Empirical Superiority of Our Method}}
In order to investigate the influence of different dynamic supervision adjustment strategies, we carefully compared our approach with existing competitive methods in Table~\ref{dynamic_weight}, including \textit{1)} ``\textbf{ViLT}'' 12 layers ViLT~\cite{kim2021vilt} as a teacher to provide soft labels and weights, \textit{2)} ``\textbf{Directly fine-tuning}'' directly finetuning 6 layers ViLT using downstream loss without distillation, \textit{3)} ``\textbf{Vanilla KD}'' 6 layers ViLT distilled by 1), \textit{4)} ``\textbf{WSL KD}'' is an existing strong baseline -- weighted soft labels KD~\cite{DBLP:conf/iclr/ZhouSCZWYZ21/wsl}, which dynamically weights the sample level by combining elements like teacher and student losses, as well as the training step, and \textit{5)} ``\textbf{Student-Uncertainty}'' follows our framework but supervised with the student uncertainty rather than teacher.

Clearly, ``vanilla KD'' improves the image-text retrieval results by averaged 7.1 points compared to directly fine-tuning the 6 layers ViLT without KD, i.e. ``Directly fine-tuning'', proving the effectiveness of knowledge distillation. Going a step further, the WSL weighting method ``WSL KD'' that combines multiple factors can push the effects of distillation to a significantly better level, i.e. averaged 1.9 points of improvements against the static ``vanilla KD''.

Surprisingly, we discovered that using student uncertainty as a weight for dynamic supervision caused the model to fail in convergence, denoted by ``N/A''.
One possible reason is that students' optimization directions may be incorrect. And such incorrect supervisions are propagated to the students' learning process, exacerbate the errors, and eventually make the networks collapse.

``DCD'' that employs the uncertain information from the knowledge-rich teachers to obtain dynamically weighted signals, by contrast, makes the training process stable, leading to further improvements. Compared with the ``Vanilla KD'' and competitive ``WLS KD'', DCD brings an average R@1 improvement of 4 and 2 points, respectively, validating the superiority of our approach.

\paragraph{\textbf{Where Do the Improvements Come From}}
In order to more intuitively show where the improvements come from, we visualize the learning dynamics of the student network in terms of the matching score of true positives and true negatives. 
Figure~\ref{fig:fig4} depicts the matching scores of positive pair and top-3 negative pairs, including the matching score of \textit{a)} the positive pair, \textit{b)} the most similar negative pair, \textit{c)} the second similar negative pair,
and \textit{d)} the third similar negative pair. 
When performing image-text retrieval tasks, we normally expect that matched image-text pairings have higher matching scores while dissimilar pairs have lower matching scores. Namely, a well-trained model is expected to have a high degree of separability between positives and negatives.

Overall, compared to the static ``Student w/ vanilla KD'' and dynamic supervision method ``Student w/ WSL KD'', our method (DCD) ``Student w/ Ours'' obtains \textbf{a higher degree of separability between positives and negatives}.

In particular, we show this with the matching score of the \textsc{positive} pair in Figure~\ref{4a}. Although our method is slightly lower than students with vanilla and WSL KD, the difference is not significant, or even comparable.

However, as for \textsc{negative} samples, our method significantly reduces the matching score of negative examples, that is, our method could distinguish the true negatives better than other students, as shown in Figure~\ref{4b}, \ref{4c} and \ref{4d}. 
Matching score observations on true positives and true negatives demonstrate that \textit{our dynamic supervision empowers students with a higher degree of separability between positives and negatives}, thus leading to improvements.

\subsubsection{\textbf{Ablation Studies}}
\label{subsubsec:ablation}
To demonstrate the effectiveness of our dynamic contrastive distillation framework, a comprehensive component wise ablation analysis is performed. The results are reported in Table~\ref{components_f} and \ref{components_ms}. Here, we use the vanilla KD described in Section \ref{section:kd} as the baseline. On both datasets, with our data selection and knowledge adjustment (``DS\&KA''), we get an average improvement of almost 2 points beyond the baseline. Also, it can be seen that, based on the strategy ``DS\&KA'', our soft-label weighting (``SW'') and hard-label weighting (``HW'') further obtain consistent improvements on both datasets. 
In particular, our ``HW'' gets more improvement on R@1, with averages of 1.9 and 0.8 points on Flickr30K and MS-COCO. 
Finally, combining all components ``FULL'', we achieve a further improvement, on average.

\subsubsection{\textbf{Generalizability of the Dynamic Distillation Framework}}
\label{subsubsec:universality}
To verify our framework as a plug-and-play component applicable to other vision-language pretraining (VLP) models that are based on modal interaction fusion, we conduct experiments upon a current SOTA VLP model METER~\cite{dou2022an}. 
METER is a dual-stream architecture that performs training of image-text pair similarity by means of a heavier modal encoder to obtain the encoding of the respective modalities and a heavier modal interaction layer for fusion encoding. Using our framework, its fused modal encoding (co-attention layers \cite{vilbert}) can be compressed, and the results are shown in Table \ref{Generalizability}.  We compress the co-attention layer \cite{vilbert} of METER from 6 layers to 3 layers and report the retrieval performance on Flickr30K with fine-tuning only, vanilla KD, WSL \cite{DBLP:conf/iclr/ZhouSCZWYZ21/wsl} weighting, and our distillation framework. The settings are consistent with the original paper~\cite{dou2022an}.
The results show that our distillation framework works well on compression co-attention, demonstrating its universality.

\section{Conclusion}
In this paper, we proposed a play-and-plug dynamic contrastive distillation framework named DCD, which consists of two major aspects, dynamic data selection and dynamic supervision adjustment, for the image-text retrieval task.
Extensive experiments upon different VLP models (ViLT and METER) demonstrate that dynamic adjustments in both data and supervision according to teachers' uncertainty estimation can effectively improve student performance and learning efficiency.
Further analyses reveal that the improvement comes from 1) fully mining the hard negative samples, and 2) providing a higher degree of separability between positives and negatives. We hope that our method could shed light on more image-text tasks in the future.

\section*{Acknowledgment}
This research was funded by National Natural Science Foundation (No.61902093), Natural Science Foundation of Guangdong (No.2020A1515010652), Shenzhen Foundational Research Funding (No.20200805173048001),  PINGAN-HITsz Intelligence Finance Research Center, Science and Technology Innovation 2030 –“Brain Science and Brain-like Research” Major Project (No. 2021ZD0201405).

\ifCLASSOPTIONcaptionsoff
  \newpage
\fi

\bibliographystyle{IEEEtran}
\bibliography{arxiv_version}
\end{document}